\documentclass[twocolumn,showpacs,floatfix,superscriptaddress,prl]{revtex4}

\usepackage{graphicx}
\usepackage{subfigure}

\begin{document}

\preprint{preprint}

\title{Critical Field Strength in an Electroclinic Liquid Crystal Elastomer
}
\author{Christopher M. Spillmann}
\email{christopher.spillmann@nrl.navy.mil}
\affiliation{
Center for Bio/Molecular Science and Engineering, Code 6900, Naval Research Laboratory, 4555 Overlook Ave. SW, Washington, D.C., 22375 USA
}
\author{Amit V. Kapur}
\affiliation{
Center for Bio/Molecular Science and Engineering, Code 6900, Naval Research Laboratory, 4555 Overlook Ave. SW, Washington, D.C., 22375 USA
}
\author{Frank W. Bentrem}
\affiliation{
Center for Bio/Molecular Science and Engineering, Code 6900, Naval Research Laboratory, 4555 Overlook Ave. SW, Washington, D.C., 22375 USA
}
\affiliation{
Marine Geosciences Division, Naval Research Laboratory, Stennis Space Center, Mississippi, 39529 USA} 

\author{Jawad Naciri}
\affiliation{
Center for Bio/Molecular Science and Engineering, Code 6900, Naval Research Laboratory, 4555 Overlook Ave. SW, Washington, D.C., 22375 USA
} 

\author{Banahalli R. Ratna}
\affiliation{
Center for Bio/Molecular Science and Engineering, Code 6900, Naval Research Laboratory, 4555 Overlook Ave. SW, Washington, D.C., 22375 USA
}

\date{\today}

\begin{abstract}

We elucidate the polymer dynamics of a liquid crystal elastomer based on the time-dependent response of the pendent liquid crystal mesogens. The molecular tilt and switching time of mesogens are analyzed as a function of temperature and cross-linking density upon application of an electric field. We observe an unexpected maximum in the switching time of the liquid crystal mesogens at intermediate field strength. Analysis of the molecular tilt over multiple time regimes correlates the maximum response time with a transition to entangled polymer dynamics at a critical field strength.

\end{abstract}

\maketitle

%\section{Introduction}
%\label{sec:intro}

 In the 1970s, Garoff and Meyer were the first to describe a distinctive class of liquid crystal (LC) that exhibits a chiral smectic-$A$ phase (Sm-$A^*$) consisting of chiral molecules with a permanent dipole close to the chiral center \cite{garoff77}. Application of an electric field couples the molecular dipole to the field and results in a molecular tilt ($\theta$) in a plane orthogonal to the transverse component of the dipole. This response was termed the electroclinic effect or soft mode, analogous to the softening of the vibration mode in a ferroelectric material as it is cooled to the Curie temperature \cite{garoff77}, and has been well documented with several LC moieties \cite{abdulhalim91,andersson88,garoff79,nishiyama87}.

\begin{figure}[b]
%\subfigure[]{
%\includegraphics[width=3.325in]{fig1b.eps}
%}
%\subfigure[]{
\includegraphics[width=3.375in]{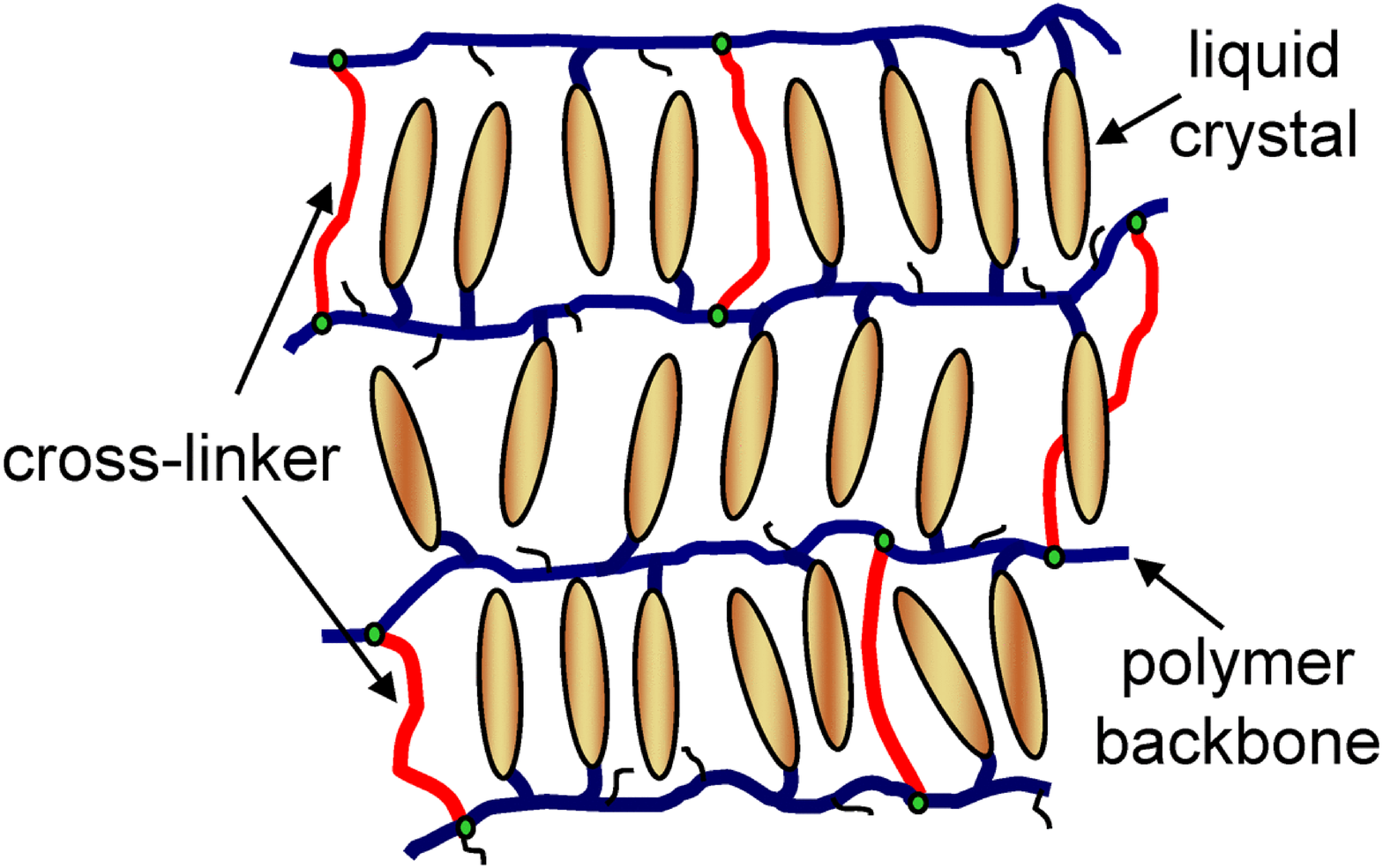}
%}
\caption{\label{fig:1} Schematic representation of smectic LC mesogens in an elastomeric network.}
\end{figure}

The reorientation of the molecular dipoles and the concomitant tilt in response to an applied electric field is resisted by the intrinsic viscosity of the material. If the LC molecules are now tethered to a polymer backbone (Fig.~\ref{fig:1}), the electroclinic response to an electric field is significantly constrained. The molecules maintain much of the rotational freedom necessary for the permanent electric dipoles to align with the applied field, but the tilt response is significantly retarded by the entangled polymer backbone. The significant influence of the polymer network was first noted by Gebhard and Zentel when they studied an elastomeric ferroelectric LC system \cite{gebhard99,gebhard00a}. They noted that the polymer network ``counterforce'' could be represented as a spring resisting the molecular tilt in the Sm-$A^*$ phase \cite{gebhard99}. However, the dynamic nature of the interplay between the LC mesogens and the underlying polymer network remains largely unexplored.
 
In this Letter, we investigate the polymer dynamics of an electroclinic LC elastomer as a function of cross-linking density, temperature, and field strength. Since the LC molecules are coupled to the polymer network, the molecular tilt response is used as an indirect probe to elucidate the polymer dynamics in the presence of an applied field. We observe multiple time domains in the tilt response and also a critical field strength corresponding to the maximum switching time of the tethered LC mesogens. At this critical field strength, the response time shows power-law dependence and supports the presence of entanglement dynamics in the elastomer system at higher field strengths. Thus, we demonstrate the ability to elucidate the polymer dynamics of the system based on the reorientation of the pendent LC elements.

Based on Landau theory \cite{clark91}, the electroclinic response of a monomeric system to an applied field, $E$, at temperatures sufficiently above the Sm-$A^*-$Sm-$C^*$ (chiral smectic-$C$) phase transition results in a linear response of $\theta$ to $E$ and a characteristic switching time $\tau$ independent of the applied field. As the system is cooled and approaches the Sm-$A^*-$Sm-$C^*$ transition, both $\theta$ and $\tau$ increase at a given field strength. The switching time also shows a dependence on $E$ such that it monotonically decreases with increasing field strength. It has been previously demonstrated that the presence of a Sm-$A^*-$Sm-$C^*$ transition is not a necessary condition for the electroclinic effect \cite{crawford95,shashidhar00}. This holds true for the $\theta$ and $\tau$ response of the monomeric mixture used in the current study (see supplementary material, \cite{supplementary}).

Engineered coupling of electroclinic LCs to a polymer network allows for an applied field to change the orientation of the pendent LC molecules and subsequently alter the elastomer system in a reversible manner \cite{hiraoka09,kohler05,spillmann07b,stannarius02}. We have developed a free-standing electroclinic elastomer and examined the molecular tilt, macroscopic actuation, and molecular packing of the system \cite{spillmann07b,spillmann08}. We now examine the time-dependent molecular response and for the first time use this information to shed light on the polymer dynamics.

%\section{Experimental Part}

%\subsection{Materials and Sample Preparation}

Synthesis of the polymerizable LC components, the diacrylate cross-linker, and preparation of the elastomer has been previously reported \cite{spillmann07b,artal01}. The structure of the molecular components and details of the sample preparation are provided as Supplementary Material. Three samples were prepared with 0, 2.5, and 5 mole percent (mol\%) of the cross-linker in EHC cells (E.H.C. Co., Ltd., Tokyo, Japan) to provide known sample thickness and transparent indium tin oxide (ITO) electrodes for electric field application. Electro-optic tilt angle and response time measurements of the electroclinic samples follow the example of Lee and Patel \cite{lee89,lee89b} and are captured under temperature-controlled conditions with a polarized light microscope. This approach is sensitive to changes in the orientation of the rigid core (phenyl rings) of the electroclinic molecules with respect to the crossed polarizers. The tilt angle $\theta$ of the LC mesogens measured upon application of an electric field is given by
\begin{equation}
\theta = \frac{1}{4} \sin^{-1} \left(\frac{\Delta I}{I_{\text{max}}-I_{\text{min}}}\right),
\end{equation}
where $I_{\text{min}}$ and $I_{\text{max}}$ are the minimum and maximum transmitted light intensity with the LC molecular director positioned 0 and $\pi/4$ radians with respect to the polarizer, respectively. With the sample positioned $\pi/8$ radians with respect to the polarizer, a bipolar square wave ($+V$ to $-V$) was applied at incrementing field strengths and the change in transmitted light intensity through the sample, $\Delta I$, was measured. The response time of the molecular tilt is defined as the time required for the molecular tilt to change from 10\% to 90\% of the full tilt angle upon reversal of the electric-field polarity.
 
%\section{Results and Discussion}

\begin{figure}
\includegraphics[width=3.375in]{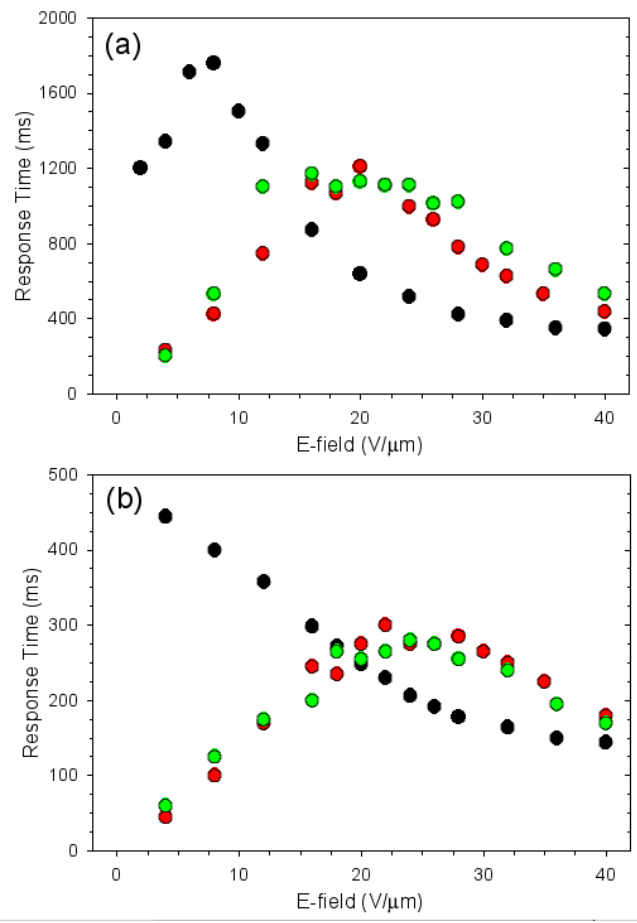}
\caption{\label{fig:3}Response time of electroclinic polymer (black circles) and elastomer with 2.5 (red circles) or 5 mol\% (green circles) cross-linking at (a) $35^\circ$C and (b) $40^\circ$C.}
\end{figure}

The tilt angle of the electroclinic polymer and elastomer samples showed characteristic curves that decreases with increasing temperature (Supplementary Material, Fig.~S4). However, the response time showed an unexpected initial {\it increase} with increasing field strength, as shown in Fig.~\ref{fig:3}. The response time of elastomer samples reached a maximum at intermediate field strengths of 15--25 V/$\mu$m and then began to decrease as $E$ was further increased. The average field strength at which the maximum $\tau$ was observed in the elastomer did not change significantly as a function of temperature. In addition, no significant difference in $\tau$ or the field strength at the maximum response time was observed when the cross-linking was either 2.5 or 5 mol\%. The polymer follows a notably different trend. At $35^\circ$C, the peak response time was observed at 8 V/$\mu$m [Fig.~\ref{fig:3}(a), black circles] and increasing the temperature to $40^\circ$C altered the response to closely resemble the shape expected for a monomeric sample [Fig.~\ref{fig:3}(b), black circles]. At lower temperatures approaching the glass transition of $\sim$28$^\circ$C, the effective polymer entanglement is large, increasing the resistance to molecular tilting with in a pronounced increase in $\tau$. These molecular observations provide critical insight of the elastomer response of an electroclinic system to applied fields and expose the presence of two opposing forces in the material: the force generated by the reorientation of the LC mesogens in response to an applied field and the resistive force of the polymer network.

\begin{figure}[b]
\includegraphics[width=3.375in]{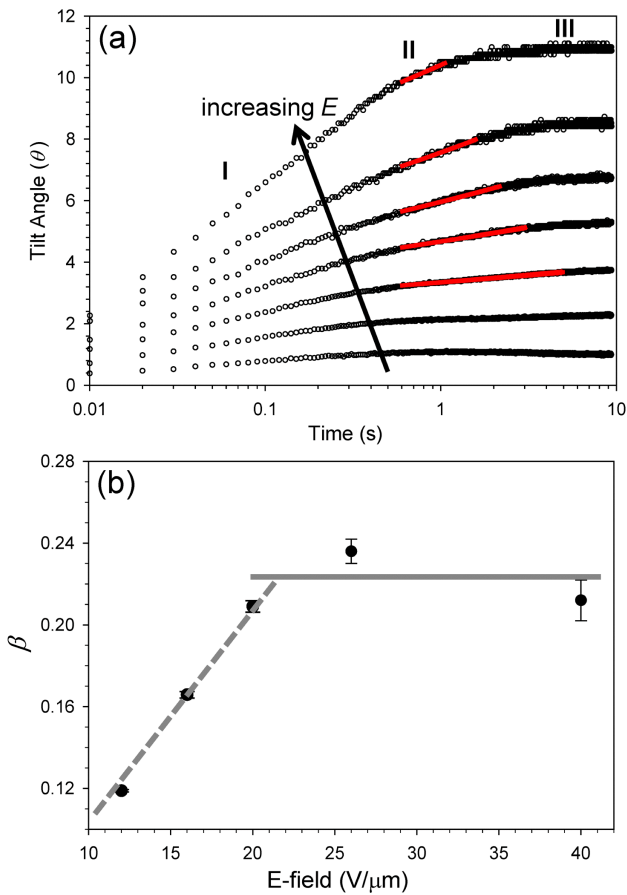}
\caption{\label{fig:4} (a) Semi-log plot of $\theta$ versus time of elastomer with 5 mol\% cross-linking at $35^\circ$C. Domains of the response are identified as regions I, II, and III. Field strengths are 4, 8, 12, 16, 20, 26, and 40 V$/\mu$m. (b) $\beta$ plotted as a function of field strength. Guide lines are overlaid for $10 \le E \le 20$ and $20 \le E \le 40$.}
\end{figure}

The origin of the light intensity changes used to monitor $\theta$ and $\tau$ is the reorientation of the LC mesogens in response to an electric field. Since these molecules are coupled to the polymer network, we explored the possibility of using this information to probe the polymer dynamics of the system by examining the time dependent response over a range of field strengths. The use of the time-dependent tilt angle as an analysis tool for the polymer dynamics of the system is analogous to a previously reported method that utilized light scattering of nanocolloidal probes adsorbed to a polymer matrix \cite{sprakel07}. In our system, it is the coupling of the LC mesogens to the network that serves as a sensitive measurement of the polymer dynamics. The dynamics of a polymer system is defined by the relationship of the mean square displacement of individual segments of a polymer chain, $\langle r(t)^2 \rangle$, to time $t$, as
\begin{equation}
\langle r(t)^2 \rangle = at^\beta,
\end{equation}							
where $a$ and $\beta$ are the creep coefficient and exponent, respectively \cite{degennes79,mcleish02}. Since the mesogens are coupled to the polymer backbone and $\theta$ is a measure of the mesogen displacement, i.e., tilt, in response to an applied field, we assume $\theta(t)$ is an indicator, and, therefore, approximately equivalent to $r(t)$ in the current system. This indirect measure of the polymer dynamics is most relevant at higher field strengths, where the coupling of the polymer backbone to the mesogenic response is realized.

Logarithmic analysis of the relationship between $\theta$ and time is used to examine the polymer dynamics of the electroclinic elastomer as a function of $E$. Figure \ref{fig:4}(a) shows a semilog plot of $\theta$ developing as a function of time at several electric field strengths upon switching the field polarity for an elastomer sample. The tilt angle advances through three time regimes (regions I, II, and III) and is described in terms of the interplay between the polymer backbone and pendent LC mesogens. Region I is the initial uncoupled response
where the LC mesogens rapidly tilt with little resistance from the polymer backbone. Region II is where the tilt response of the coupled mesogens becomes significantly influenced by the underlying polymer network prior to saturation of the tilt angle in region III. 

The molecular tilt in region I follows a logarithmic relationship during the initial response to the e-field. Region II was defined as the intermediate regime between the initial logarithmic response and the saturated tilt angle estimated from linear fits to region I and III in Fig.~\ref{fig:4}(a). A log−log analysis of region II shows evidence of power-law dependence. Linear regression [Fig.~\ref{fig:4}(a), red lines] provides detail about the creep exponent $\beta$, which is expected to depend on field strength, temperature, and cross-linking. The values of $\beta$ as a function of electric field strength are shown in Fig.~\ref{fig:4}(b). We note that this analysis excluded the lowest field strengths ($<10$ V/$\mu$m), where the coupling between the mesogenic tilt and polymer backbone accommodates small tilt angles without a significant influence from the polymer backbone. At field strengths ranging from $\sim$10--20 V/$\mu$m, there is a linear dependence on the applied field. At field strengths greater than 20 V/$\mu$m, $\beta$ saturates to a constant value of $\sim$0.22. Various models of polymer systems predict power-law displacement, including Rouse, reptation, and entanglement dynamics \cite{degennes79,mcleish02}. The saturating value of $\beta$ strongly suggests a $t^{1/4}$ dependence, which indicates the polymer backbone may be constrained by the dynamics of polymer entanglement in this time regime \cite{degennes79,mcleish02}. It is important to note that previous comparative study on entangled and covalently cross-linked polymers had shown that the dynamics are indentical for the two cases \cite{nicolai99}. The transition to entanglement dynamics at $\sim$20 V/$\mu$m occurs at the same field strength that, on average, has the longest response time in the elastomer samples (see Fig.~\ref{fig:3}). Therefore, the polymer dynamic analysis supports the notion of a critical field strength at which the underlying polymer network begins to significantly reorient in response to the external stimulus.

The presence of a maximum $\tau$ at an intermediate $E$ quantifies the ability of the electrically-induced mesogenic tilt to affect the reorientation of the polymer backbone at various field strengths. At the lowest applied fields, the mesogenic tilt is only a few degrees and the stress on the polymer backbone is negligible and $\tau$ relatively fast. At slightly higher field strengths, the molecular tilt begins to impose a significant tug on the polymer backbone. The restraint or creep of the polymer backbone is observed as an increase in $\tau$. The force applied by the LC reorientation has increased sufficiently to induce slow realignment of the polymer backbone from its equilibrium state. This continues as the field strength and molecular tilt increase to the point where $\tau$ reaches a maximum and the force of the mesogenic tilt and the polymer resistive force appear to reach a balance. At the highest field strengths, the force produced by the LC mesogens overcomes the resistance of the polymer backbone and the maximum tilt angle is reached at progressively faster response times. It should be pointed out that when a cross-linking agent is present in the system, there is additional restoring force introduced above and beyond the presence of the polymer backbone that accounts for the persistent critical field strength observed in the elastomer samples. A schematic representation of the polymer realignment and mechanical shear induced by the LC mesogenic tilt at increasing field strengths is provided in the Supplementary Information, Fig.~S5.
 
The molecular tilt angle and response time of an electroclinic liquid crystal have been examined as a monomeric system, a polymer, and a cross-linked network. Following polymerization, the magnitude of the molecular tilt angle remains the same with significant increases in both the response time and the field strength required to elicit a response. The presence of a critical intermediary field strength is identified where the characteristic response time reaches a maximum. The mesogenic response as a function of applied field is related to the interplay between the force arising from the molecular tilt of the liquid crystal mesogens and the resistive force of the entangled polymer backbone.  Analysis of the time-dependent response over the range of field strengths provides a novel approach to examine the polymer dynamics of the system and suggests an entanglement regime above the critical field strength.

This work was supported by the U. S. Office of Naval Research. F.W.B. also received support from the Naval Research Laboratory Advanced Graduate Research Program.

\bibliography{skb09}

\end{document}